\documentclass[12pt,a4paper]{article}

\usepackage{amsmath}
\usepackage{amsfonts}
\usepackage{amssymb}
\usepackage{amsthm}

\input{epsf}             
\def\epsfcenter#1{{\vcenter{\hbox{\epsfbox{#1}}}}} 

\newcommand{\R}{\mathbb R}              

\newcommand{\SU}{{\rm SU}}

\newcommand{\Space}{\mathcal S}  

\newcommand{\dd}{{\rm d}} 

\newcommand{\Atop}[2]{\genfrac{}{}{0pt}{}{#1}{#2}} 

\theoremstyle{plain} 

\theoremstyle{definition}

\newtheorem{example}{Example}

\theoremstyle{remark}

\begin{document}

\title{Feynman loops and three-dimensional quantum gravity}

\author{John W. Barrett
\thanks{This is based on lectures given at the Quantum Hyperbolic Geometry Workshop, Albert-Einstein-Institute June 2004, and the  Workshop on Quantum Gravity and Noncommutative Geometry, Universidade Lus\'ofona July 2004. Copyright \copyright\ John W. Barrett 2004 2005}
\\ \\
School of Mathematical Sciences\\
University of Nottingham\\
University Park\\
Nottingham NG7 2RD, UK\\
\\
john.barrett@nottingham.ac.uk}

\date{14th January 2005}

\maketitle

\begin{abstract}  This paper explores the idea that within the framework of three-dimensional quantum gravity one can extend the notion of Feynman diagram to include the coupling of the particles in the diagram with quantum gravity.
The paper concentrates on the non-trivial part of the gravitational response, which is to the large momenta propagating around a closed loop. By taking a limiting case one can give a simple geometric description of this gravitational response. This is calculated in detail for the example of a closed Feynman loop in the form of a trefoil knot. The results show that when the magnitude of the momentum passes a certain threshold value, non-trivial gravitational configurations of the knot play an important role.

The calculations also provide some new information about a limit of the
coloured Jones polynomial which may be of independent mathematical interest.
\end{abstract}

 
\section{Introduction}

Quantum gravity in 3+0 dimensions with a positive cosmological constant has a complete mathematical definition.   Therefore the focus of work on this theory lies in trying to understand the physical properties of the theory. The word `physical' here means 3+0-dimensional physics, that is, physics where there are three space dimensions and no time dimensions. Obviously this isn't true 3+1-dimensional physics which could be applied to our universe, but physics in this parallel 3+0-dimensional world.  

The intuitive idea for the definition of quantum gravity is to calculate an average of the exponential of the Einstein-Hilbert action $S$ 
\begin{equation}\label{functionalintegral}Z(M)=\int e^{iS(M,g)}\end{equation}
 over all 3-metrics $g$ on a given 3-manifold $M$. The definition of this functional integral was addressed by Ponzano and Regge \cite{PR} and Witten \cite{W1,W2}, but neither give finite definitions in general and both seem hard to make precise. The problem, as we now understand it, was that the former paper did not include a cosmological constant and the latter used lorentzian signature metrics. 
 
 Both of these defects were cured with the definition of Turaev and Viro \cite{TV}. This gave a complete definition of the functional integral for Euclidean signature metrics with a positive cosmological constant.  There is a theory defined for each integer $r\ge3$; this number fixes the cosmological constant, $\Lambda=1/r^2$. So for a fixed value of the cosmological constant, the Turaev-Viro theory determines a number for each compact 3-manifold $M$.

\subsection{Feynman diagrams}
 In \cite{B} I gave a prescription for calculating observables with this functional integral. The physical content of the theory is vastly increased with the inclusion of these observables as one can do local, rather than global, calculations. An observable is specified by a graph embedded in $M$. For each edge of the graph one can specify the value of either a distance variable or a mass variable for the edge. Then one calculates the functional integral with this observable. This is defined by including appropriate extra terms in (\ref{functionalintegral}): a delta function restricting the metrics which are summed over for the distance observable, or an additional matter source term concentrated on the graph for the mass observable. A precise definition for a particular case is given in the next section. 

The physical picture is of a system of particles moving along the edges of the graph and interacting at the vertices; the partition function calculates a quantum amplitude for this configuration.\footnote{Other proposals for including particles, including spin and particles with free ends, are given in \cite{FL1,M}.}  
The distance variable on an edge specifies a distance between the two endpoints, whereas the mass variable specifies the mass (i.e., the magnitude of the momentum) of the particle moving along the edge.
For a classical geometry, the mass variable would determine the conical defect angle of the geometry at this edge. In quantum gravity, these two variables are conjugate variables: one cannot specify the values of both. In fact the observable for one variable is a Fourier transform of the observable for the other variable. 

It seems that an appropriate way of thinking about this picture is that one is calculating the partition function for the gravitational response to a Feynman diagram. In a quantum field theory one would also have additional propagator and vertex factors for each diagram, depending on the matter fields and their interactions. These factors are not included in the purely gravitational part described here.

The idea that the particle is a quantum mechanical particle, and not just a classical source term, requires explanation.  In quantum mechanics one integrates over all possible trajectories for a particle on a fixed background metric. In quantum gravity one no longer need do this as the integration over all metrics effectively incorporates fluctuations of the trajectory of the particle\footnote{This idea goes back to Witten, at least \cite{W2}.}. The idea is that if a particle trajectory is changed by a fluctuation, the new trajectory can be mapped back to the original one by a (piecewise-linear) homeomorphism, at the expense of replacing the metric by its image under the homeomorphism. However in quantum gravity one integrates over all metrics, so this makes no difference.
 This means that calculations of the quantum gravity functional integral with external particle sources already incorporate quantum mechanical aspects of the propagation of the particles.   One can specify a fixed trajectory for the particle, obtaining the same result for any trajectory in the same knot class.

\section{A closed loop}
 In this paper I concentrate on the case where $M=S^3$ and the graph is simply a knot $K$ with a number $n$ of vertices along it. The mass variable is a half-integer $l$, $0\le l\le (r-2)/2$, and there are $n$ of these variables, one for each segment of the knot. 
The functional integral for this case is written $Z(S^3,(K,n))$.
 
 $$ \epsfbox{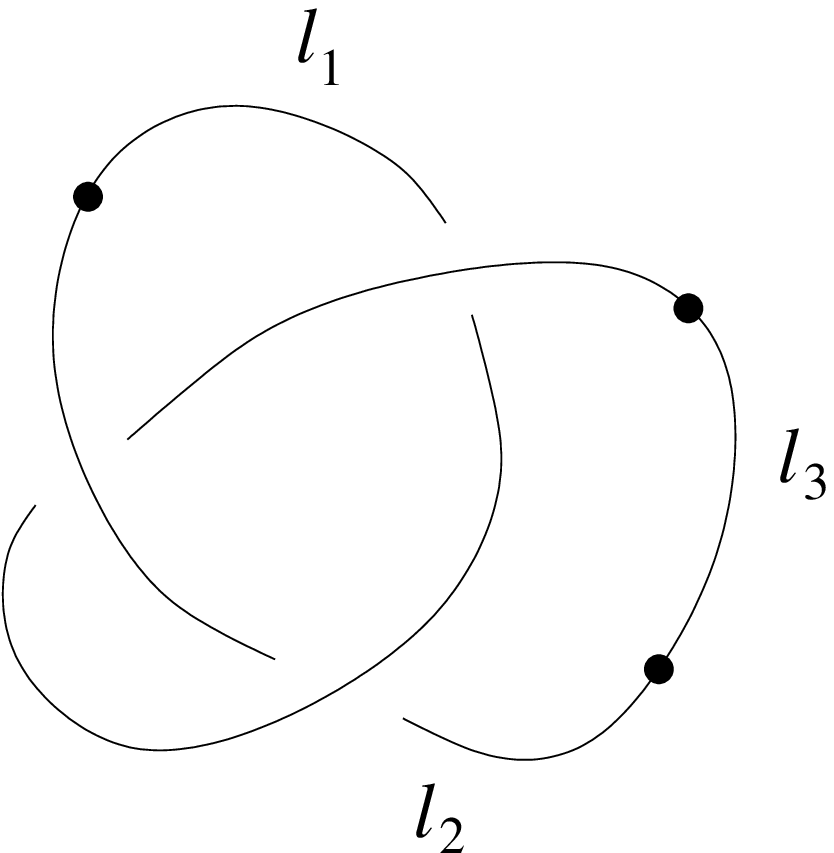}$$ 

 According to \cite{B} and \cite{BGM} the definition of $Z$ for this case can be written in terms of the coloured Jones polynomial of $K$, $J_K(l)$, where $l$ determines the spin $l$ representation of $U_qsl2$ (with classical dimension $2l+1$) and $q=e^{i\pi/r}$, using the convention\footnote{Unfortunately a number of authors write $q^{1/2}=e^{i\pi/r}$.} of \cite{KL} for $q$. This is a Laurent polynomial in $q$ and thus a function of $r$. The dependence on $r$ of this and similar formulae is generally omitted in notation such as $J_K(l)$ to avoid clutter.
 
The definition of $Z$ reduces to
 $$Z(S^3,(K,n))=Z(S^3)\frac{|J_K(l)|^2}{J_O(l)^n}\delta_{l_1l}\delta_{l_2l}\ldots\delta_{l_nl},$$
 where $O$ denotes the unknot. The first factor on the right is the normalisation coefficient
 $$Z(S^3)=\frac{2\sin^2(\pi/r)}r,$$ 
 and $$J_O(l)=(-1)^{2l}\frac{\sin\frac\pi r(2l+1)}{\sin\frac\pi r}=(-1)^{2l+1}\frac{q^{2l+1}-q^{-2l-1}}{q-q^{-1}}.$$
 Note that $Z$ is independent of the framing of the knot.

  The heuristic arguments in \cite{B} indicate that effect of the observable is a particle of mass $l+\frac12$ moving in a closed circuit around the knot $K$. The purpose of this paper is to provide substance for this argument by computing a suitable limiting case where $l \to\infty$ and $r\to\infty$, and showing that the functional integral reduces to a simple form which is related to configurations of gravitational fields with a mass proportional to $l+\frac12$ on the knot. In addition to having the simple geometrical form, the use of the limit also allows a discussion of the behaviour of the partition function for large masses (i.e., large momenta), in particular whether the partition function itself is `large'. In the Turaev-Viro model itself this is hard to formulate; not only are the masses bounded, but also the partition function is itself always finite. However in the limit which is analysed, these quantities can diverge at differing rates, so that `large' has a precise meaning.
  
The vertices on the knot merely affect the normalisation factor. Physically one can think of the vertices as marked points in the space-time through which the particle passes. In the more general framework of \cite{B}, other particles could interact at these vertices, forming a larger Feynman diagram in the space-time. In fact $Z(S^3,(K,n))$ is exactly equal to the partition function of any graph containing $K$ with the mass parameters of all the other edges set equal to 0.
$$\epsfbox{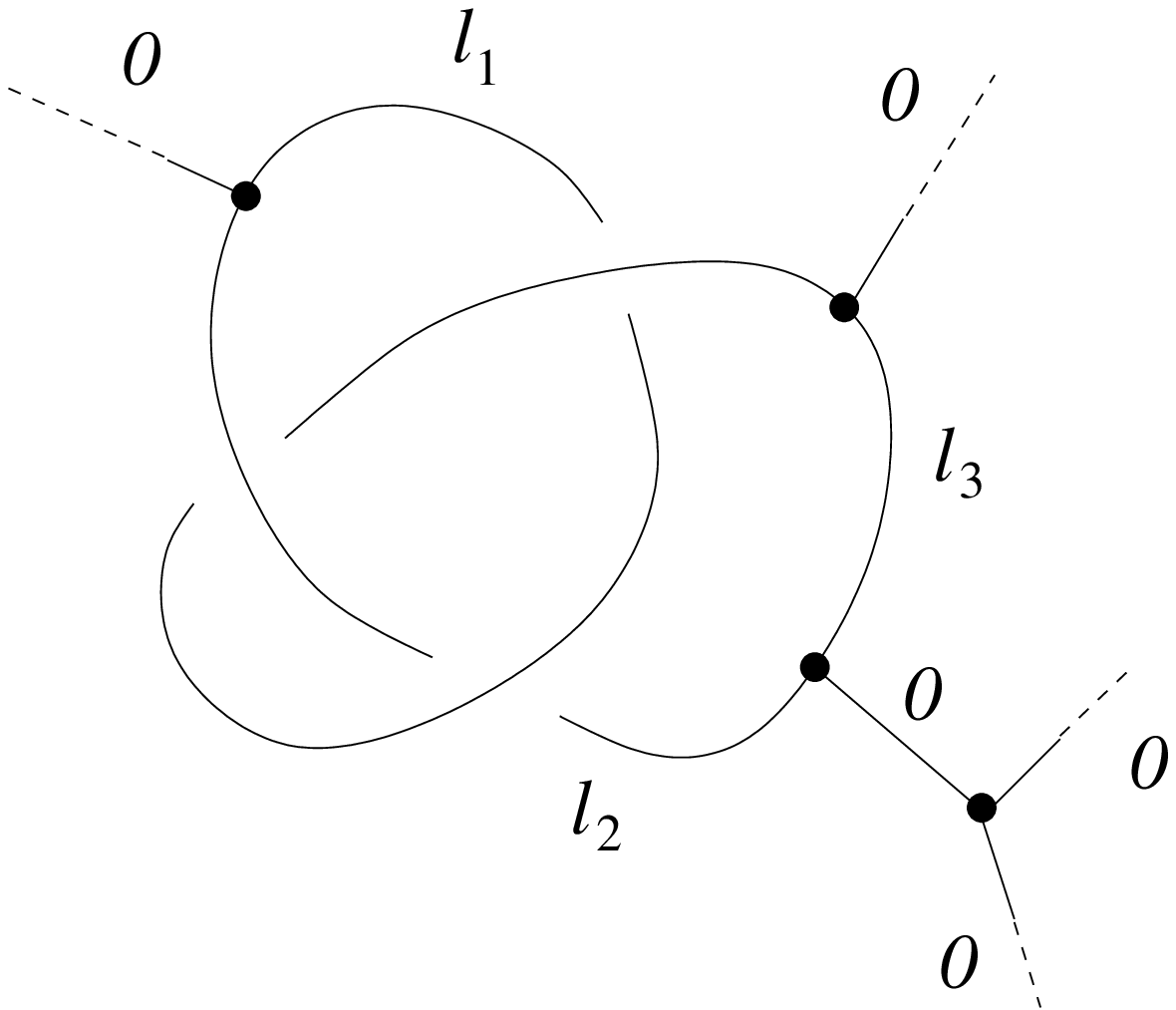}$$
Thus this paper can be viewed as an analysis of the quantum gravity effects of large momenta in a closed loop in a Feynman diagram.

  The calculation also provides some new information about limits of the coloured Jones polynomial which may be of independent mathematical interest.

 \section{The limit}
 Fix a knot $K$ and a value of $n$ and consider the limit of
 $$W(K,n,l)=(-1)^{2ln}\frac{Z(S^3,(K,n))}{Z(S^3)}=\frac{|J_K(l)|^2}{|J_O(l)|^n}$$
 as 
 \begin{equation}\label{limit}r\to\infty\quad \text{such that} \quad\frac{2l+1}r\to\frac\theta{2\pi}\end{equation} 
 for some fixed $\theta$, $0\le\theta\le 2\pi$.
 
 Note that this limit is similar to that encountered in the hyperbolic volume conjecture \cite{HV}, except that the choices of $r$ and $l$ are those which are not allowed for the hyperbolic volume conjecture. 
  
 Since $J_O(l)\sim r$ in this limit, then one has to pick the right values of $n$ to get convergence. For example, for $K=O$, $W$ does not converge for $N<2$, $W=1$ for $n=2$ and $W\to0$ for $n>2$. Thus the naive expectation would be that $n\ge2$ gives convergence. However this is in general not correct. As will be shown, this only holds for sufficiently small values of $\theta$.

 The limit can be investigated in three ways.
 
 \subsection{The Melvin-Morton expansion}
 The Melvin-Morton expansion for the coloured Jones polynomial of a 0-framed knot $K$ is a power series in $l$ and $r^{-1}$ \cite{MM},  
$$ \frac{J_K(l)}{J_O(l)}= \sum_{k,m\ge0}a_{km}(2l+1)^k\left(\frac{2\pi i}r\right)^m,$$
for which $a_{km}=0$ for $k>m$. For constant values of $l$, this is the power series in $r^{-1}$ one obtains from the perturbative expansion of the Chern-Simons functional integral.
Taking the limit (\ref{limit}) of each term in the series gives $a_{kk}(i\theta)^k$ for $k=m$ and zero otherwise.
Summing the resulting limiting series gives \cite{BG}
$$\sum_{k\ge0}a_{kk}(i\theta)^k=\frac1{A_K(e^{i\theta})},$$
where $A_K(x)$ is the Alexander polynomial of the knot $K$ (with a suitable normalisation). 

The difficulty with this calculation is that it consists of manipulations of formal power series\footnote{The convergence of the Melvin-Morton series is discussed in \cite{FM}.}, which does not actually guarantee the correct limit for non-zero values of $\theta$. In fact since $A_K$ is a polynomial, the series in $\theta$ cannot converge in the complex plane past the first zero of $A_K$.

However the calculation,
applied to the  $r\to\infty$ limit (\ref{limit}) of $W$ suggests the conjecture
$$W(K,2,l)\to\frac1{|A_K(e^{i\theta})|^2}$$
for any knot $K$ and sufficiently small $\theta$.    Note that although the Alexander polynomial $A_K(x)$ is defined only up multiplication by a normalising factor of $\pm x^k$, the value of $|A_K(e^{i\theta})|$ is uniquely determined.

\subsection{Torus knots}
The $(m,p)$-torus knot, for coprime integers $m,p$, is the knot obtained by decomposing $S^3$ as the union of two solid tori, and drawing a simple closed curve on the common boundary which intersects one meridian $m$ times and the other one $p$ times. The diagram above shows the $(2,3)$-torus knot, the trefoil.
 
Kashaev and Tirkkonen \cite{KT} give an integral expression for the coloured Jones polynomial of an $(m,p)$-torus knot. It is possible to take the limit
$$r\to\infty\quad\text{such that}\quad\frac{2l+1}r=\frac\theta{2\pi},$$
with $\theta$ a rational multiple of $2\pi$, in their formula. This is done using an asymptotic expansion, exactly as in the proof of the theorem in \cite{KT}. Similar calculations appear in \cite{ROZ}. The result is a proof that
 \begin{equation}W(K,2,l)\to\frac1{|A_K(e^{i\theta})|^2}\label{reducible}\end{equation}
 for $0<\theta<\theta_0$ and $2\pi-\theta_0<\theta<2\pi$, where $\theta_0$ is the smallest value of $\theta$ for which $e^{i\theta}$ is a root of the Alexander polynomial. Note that for torus knots, the roots of the Alexander polynomial all lie on the unit circle.
For the remaining values of $\theta$, $W(K,2,l) \sim r$ and so diverges (assuming the coefficient is not zero).  This result is very surprising; it indicates that the perturbative expansion of the gravity functional integral is misleading in this limit. Since the function diverges as $r$, this suggests that increasing $n$ to $n=3$ will give a convergent result. This is true for torus knots. The actual limit is calculated in the following example.

\begin{example} The Alexander polynomial of the trefoil is $A_K(x)=x^2-x+1$ with roots at $x=e^{\pm i\frac\pi3}$. Thus $\theta_0=\pi/3$.  A calculation shows that
\begin{equation} W(K,3,l)\to \begin{cases} 0 &
 \text{$0<\theta<\pi/3$,\quad $5\pi/3<\theta<2\pi$,} \\
 \frac\pi{4\sin^3 (\theta/2)} & 
 \pi/3<\theta<5\pi/3.\end{cases}\label{irreducible}\end{equation}
 \end{example}

\subsection{$\SU(2)$ holonomy}
The third way of understanding the limit is to construct the gravitational fields of the limiting geometry and give a heuristic model for the limit of the partition function. This will be illustrated by performing the calculations for the simplest example, the trefoil knot. 

The geometry which is relevant is that of constant curvature $1/r^2$, this being the solution of the Einstein field equations in three dimensions. Thus the geometry in the $r\to\infty$ limit is locally flat. However the inclusion of the Feynman diagram involves a source for the energy-momentum tensor along the graph. The effect of this is that the spin connection is flat on the exterior of the knot, $S^3\setminus K$, but there is a holonomy $H\colon\pi_1(S^3\setminus K)\to \SU(2)$ modelled on the holonomy of a conical metric singularity.
$$\epsfcenter{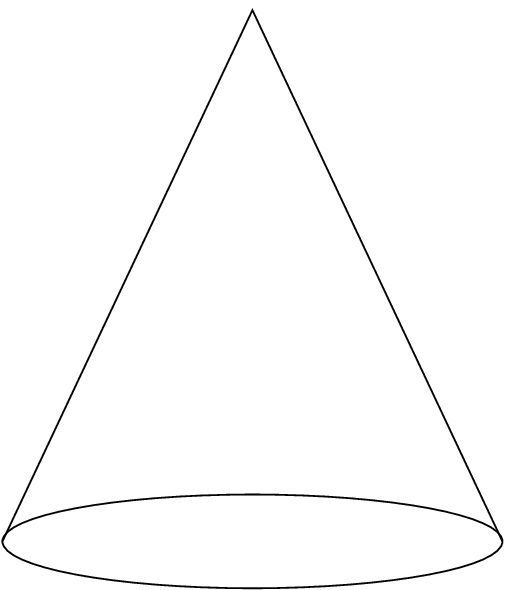}\quad\times\quad\R$$
Moreover, the conjugacy class of the holonomy around the singularity determines the mass of the energy-momentum source term on the graph. 

Therefore the model for the limit (\ref{limit}) is constructed in terms of flat $\SU(2)$ connections on $S^3\setminus K$ that have holonomy conjugate to
$$\begin{pmatrix}  e^{i{\theta/2}}&0 \\0&e^{-i{\theta/2}} \end{pmatrix}$$
for any loop which encircles the knot (i.e., bounds a disk that intersects the knot once). This idea is consistent with the observation that the functional integral for gravity without cosmological constant is one-loop exact \cite{W2}, and the 
similar notion of an observable in the Ponzano-Regge model defined in \cite{FL1,FL2}.

A flat connection on a knot can be described easily given a knot diagram. 
First, orient the knot. Then each arc on the diagram has associated to it the holonomy of the element of $\pi_1$ it generates by circulating around it in the direction of a right-handed screw. The loop is connected to a basepoint by paths which lie above the knot on the diagram. The following  diagram shows a generator of $\pi_1$ with holonomy $g$.  
$$\epsfcenter{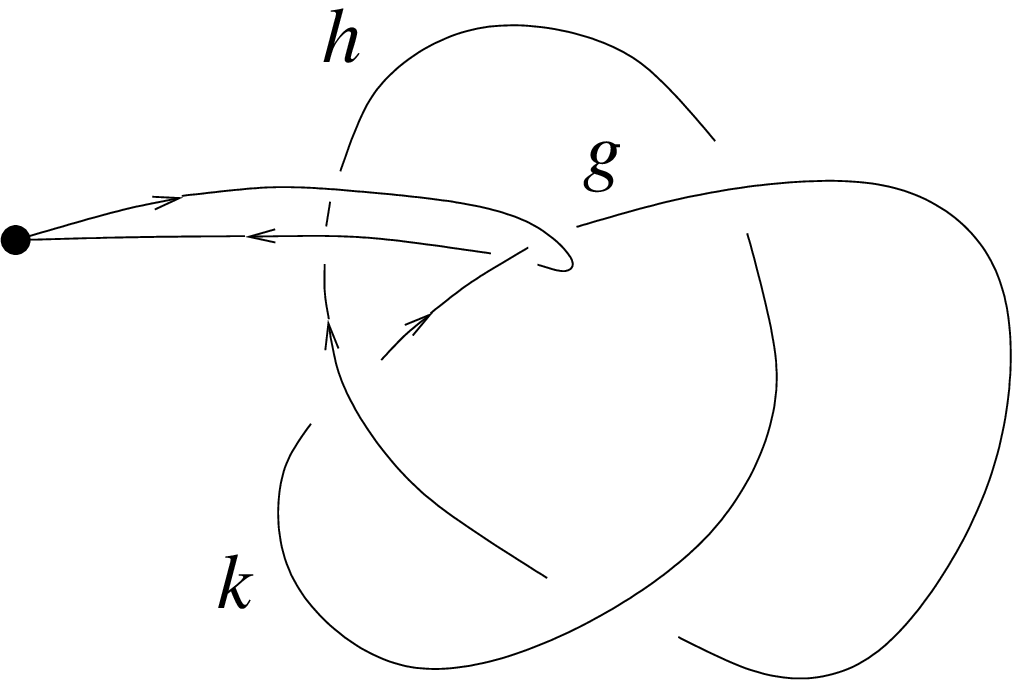} $$
At each crossing on the diagram there is a relation
$$\epsfcenter{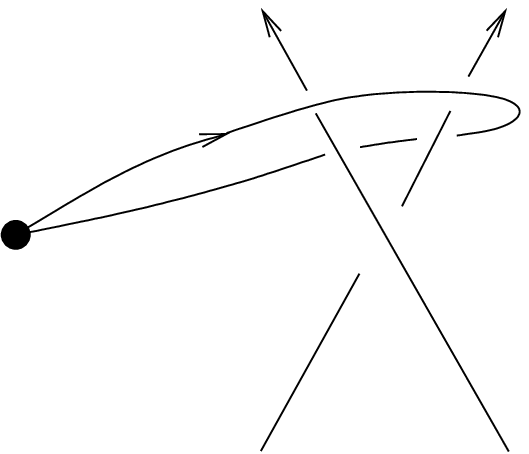}\quad=\quad\epsfcenter{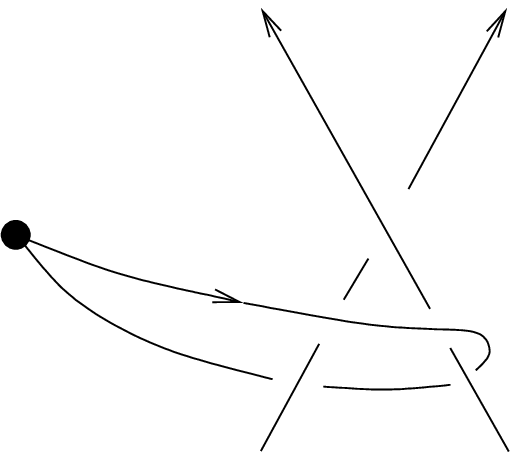}$$
For example, for the trefoil knot the holonomy is determined by $g,h,k\in\SU(2)$ with relations
$$gk=hg$$
$$hg=kh$$
$$kh=gk$$
For any knot, one of the relations is always redundant; it is implied by the others. Note that since $g,h$ and $k$ are all conjugate to each other, there is a uniquely determined conjugacy class $C_\theta$.

The set $\Space$ of solutions of these equations for the trefoil is given by the union of two sets $\Space_1$ and $\Space_2$
\begin{enumerate} 
\item $\Space_1$: $g=h=k$. This is possible for all $0\le\theta\le 2\pi$. It is a three-parameter family of solutions, with the rotations having two-dimensional symmetry orbits, and $\theta$ the remaining parameter.
\item $\Space_2$: $(hk)^3=(khk)^2=-\begin{pmatrix} 1&0\\0&1\end{pmatrix}$, $g=h^{-1}kh$. This is only possible for $\pi/3\le\theta\le5\pi/3$. It is a four-parameter family of solutions, with three-dimensional symmetry orbits and, again, $\theta$ the remaining parameter. Geometrically, the axes of the rotations corresponding to $g$, $h$ and $k$ form the vertices of an equilateral spherical triangle on $S^2$ with angles of the triangle equal to $\theta$.
\end{enumerate}
The two sets of solutions meet at the roots of the Alexander polynomial, $\theta=\pi/3$ or $5\pi/3$, as can be seen from the fact that $\pi/3$ is the minimum angle for an equilateral spherical triangle.
 
 The model for the partition function in the $r\to\infty$ limit is the integral
 $$I_K=\int_{\Atop{g\in C_\theta}{h,k\in\SU(2)}}\dd g\dd h\dd k\; \delta(hgh^{-1}k^{-1})\delta(khk^{-1}g^{-1}),$$
 with analogous expressions for any knot diagram.
 The delta functions are three-dimensional delta functions at the identity element of $\SU(2)$. The integral is constructed by enforcing all the relations in the fundamental group by including one factor for each crossing excepting one. Accordingly the corresponding sum for a finite group in place of $\SU(2)$ would count the number of flat connections with a given conjugacy class of holonomy around the knot. However replacing the finite group with the Lie group $\SU(2)$ is not trivial, and the integral may not exist. If it does exist, the expression is invariant under Reidemeister moves.
 
 For the trefoil knot, the integral splits into two parts corresponding to the two branches of solutions to the relations,
 $$I_K=I^1_K+I_K^2.$$
 The integral for the first branch, $g=h=k$ can be done. Fixing the value of $g$ and doing the integrals over $h$ and $k$ by linearising around the solution $g=h=k$ results in 
 $$I^1_K=\frac1{|D|}\int_{g\in C_\theta}\dd g,$$
 where $D$ is a $6\times6$ determinant 
 $$D=\left|\begin{matrix}1-X & X\\-1&1-X\end{matrix}\right|$$
 constructed from the $3\times 3$ matrix $X$ determined by the rotation induced by $g$. Note that the determinant of the $2\times 2$ matrix given by treating $X$ as an indeterminate is the Alexander polynomial of the knot.
Therefore, since the matrix $X$ has eigenvalues $1$, $e^{i\theta}$ and $e^{-i\theta}$, 
 $$D=\det(A_K(X))=A_K(1) A_K(e^{i\theta}) A_K(e^{-i\theta}).$$
 Thus
 $$I_K^1=c \frac{\sin^2\theta/2}{|A_K(e^{i\theta})|^2}$$
 with $c$ a numerical constant. The unknot gives
 $$I_O=I_O^1=c \sin^2\theta/2,$$
so the ratio is
$$\frac {I_K^1}{I_O}=\frac1{|A_K(e^{i\theta})|^2},$$
in agreement with (\ref{reducible}). This gives a heuristic explanation of (\ref{reducible}) whenever $\theta$ is such that the only flat connections are given by $g=h=k$. This calculation generalises to any knot and so there is a plausible conjecture that  (\ref{reducible}) holds for any knot whenever $\theta$ is such that the only $\SU(2)$ holonomy is the trivial one.

The second integral $I_K^2$ for the trefoil is more problematic. Due to the fact that the symmetry orbit is three-dimensional, it is clear that there are too many delta functions and that the integral is not defined when $\pi/3\le\theta\le5\pi/3$ as it stands. This explains the divergence of $W(K,2,l)$. It would be interesting to construct a modified integral which gives the correct limit (\ref{irreducible}) for $W(K,3,l)$.

\section{Gravitational fields}
Finally, some remarks on the gravitational fields for this heuristic limit. The $\SU(2)$ holonomy determines a flat $\SU(2)$ connection on the knot exterior. However this is only half of the data for a gravitational field. A gravitational field in the first-order formalism is given by an $\SU(2)$ connection and an $\R^3$-valued one-form, the triad. This data can be viewed as a connection for the semi-direct product $\SU(2)\times_S \R^3$, which has product 
$$(H_1,T_1)(H_2,T_2)=(H_1 H_2,H_2^{-1}(T_1)+T_2).$$
In this gauge theory perspective, it is possible for the triad to be degenerate (i.e., not an invertible linear mapping). Further conditions that would guarantee a non-degenerate metric are not considered here.

Picking a basepoint in the geometry gives a holonomy in this group for every closed loop. Therefore a gravitational field compatible with $H$ is associated to
 a lift of $H$ to a homomorphism
$$\widehat H\colon \pi_1(M\setminus K)\to  \SU(2)\times_S\R^3$$
of the form $\widehat H(\gamma)=(H(\gamma),T(\gamma))$.
Such lifts are easy to describe. The translational part $T$ is determined by a tangent vector to the space $\Space$ according to the formula
\begin{equation}\label{tangent}T(\gamma)=H(\gamma)^{-1}\frac{\dd H}{\dd t}(\gamma).\end{equation}
This follows by differentiating the condition that $H$ is a homomorphism.

Conjugation of $\widehat H$ by an element   
$E\in\R^3\times_S \SU(2)$   gives
$$\widehat H'(\gamma)=E \widehat H(\gamma) E^{-1}$$
for all $\gamma$. This corresponds to changing the basepoint of the loops in the gravitational field, and an $\SU(2)$ gauge transformation.  Thus the information about the geometry of the gravitational field is really encoded in the equivalence classes of homomorphisms $\widehat H$ modulo conjugation. 

Given $H$ and an element of the Lie algebra of $\SU(2)$, $Y\in\R^3$, one can define the one-parameter family of holonomies 
$$ \gamma\mapsto\exp(tY) H(\gamma)\exp(-tY).$$
By (\ref{tangent}), this gives
$$T(\gamma)=H(\gamma)^{-1}(Y)-Y$$
and so the gravitational field
\begin{equation}\gamma\mapsto(H(\gamma),T(\gamma))=(1,Y)(H(\gamma),0)(1,-Y),
\label{conjugate}\end{equation}
which is conjugation of the trivial solition $T(\gamma)=0$ by $E=(1,Y)$.

To understand the significance of these solutions it is necessary to introduce some more geometric considerations.
Let $\gamma_0$ be a loop which encircles the knot (i.e., just passes around one strand, as explained above). A requirement that the holonomy be modelled on the metric of a cone is that the element $\widehat H(\gamma_0)$ acts on $\R^3$ by a rotation about some axis (not necessarily through the origin), rather than the more general screw motion. This is equivalent to the requirement that there exists an element $E=(1,Z)$ (depending on $\gamma_0$) such that
 \begin{equation}\label{spinless} (H(\gamma_0),0)=E \widehat H(\gamma_0) E^{-1},\end{equation}
 the holonomy on the left being that given by a basepoint right next to the conical singularity.
 For a cone, $|Z|$ is just the distance of the basepoint of the holonomy from the nearest point on the conical singularity. From (\ref{spinless}) one can see that $T(\gamma_0)$ is orthogonal to the axis of rotation of $H(\gamma_0)$ and
 $$ |T(\gamma_0)|=2|Z|\sin\theta/2.$$
 Therefore, $|T(\gamma_0)|$ gives a measure of the distance of the basepoint from the conical singularity.
 
 \subsection{Example}
 Now these observations are applied to the example of the trefoil knot. For any $\theta$, the zero solution $T(\gamma)=0$ is always a possible gravitational field. In this case the basepoint is arbitrarily close to all arc segments of the knot simultaneously. Clearly the knot has zero size in this gravitational field.
 
 For generic $\theta$, there are no non-trivial solutions (i.e., not conjugate to the zero solution) for $T$ which satisfy the condition that $T(\gamma_0)$ is orthogonal to the axis of rotation for $H(\gamma_0)$. Therefore one does not have any more gravitational fields beyond the trivial one. This is because in the interior of either $S_1$ or $S_2$ the only curves at constant $\theta$ are rigid motions of the configurations, to which (\ref{conjugate}) applies.
 
 For the particular values $\theta=\theta_0$ or $2\pi-\theta_0$, the roots of the Alexander polynomial, there are non-trivial solutions. These are given by the tangent vectors to the branch $\Space_2$ of solutions, at the endpoint of $\Space_2$, where the triangle has zero size. Representative solutions for $T$ (one in each conjugacy class) corresponding to the three generators $g$, $h$ and $k$ are the vectors lying in the plane orthogonal to the common axis of rotation. These have equal length and at at equal angles to each other.
 $$\epsfbox{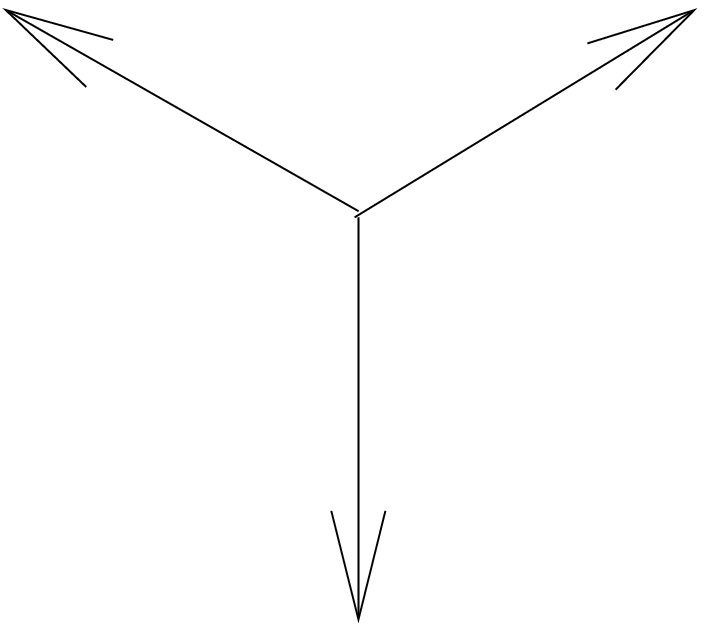}$$
 One of these vectors can be chosen to have arbitrary orientation and magnitude; this gives a two-dimensional space of the solutions.
 Conjugation by $E$ affects these by adding the same vector to all three. Thus although one can reduce one of the vectors to zero by conjugation, the other two will not be zero. Thus there is no location of the basepoint in the geometry at zero distance from all of the arcs in the knot diagram. In this case the knot has a finite non-zero size in the gravitational field. Since the overall scale is arbitrary, this size can be arbitrarily large. Thus the picture that emerges is that the qualitatively different behaviour of the limit of the coloured Jones polynomial for $\theta>\theta_0$ is due to the appearance of gravitational fields in which the knot is arbitrarily large at the critical value of the mass parameter $\theta=\theta_0$.

 \subsection*{Acknowledgments}
  Thanks are due to conversations with Jorma Louko, Jo\~ao Faria Martins, Aleksandar Mikovic, Ileana Naish-Guzman and Dylan Thurston.

\end{document}